\documentstyle[prb,aps,preprint,tighten]{revtex}
\begin{document}
\title{Does the attractive Hubbard model support larger persistent
currents than the repulsive one ?}
\author{Thierry Giamarchi}
\address{Laboratoire de Physique des Solides, Universit{\'e} Paris--Sud,
                   B{\^a}t. 510, 91405 Orsay, France\cite{junk}}
\author{B. Sriram Shastry}
\address{Department of Physics, Indian Institute of Science,
Bangalore 560 012, India\cite{junk2}}
\maketitle

\begin{abstract}
We consider a one-dimensional Hubbard model in the
presence of disorder. We compute the charge stiffness
for a mesoscopic ring, as a function  of the size $L$,
which is a measure of the permanent currents.
We find that for finite disorder the permanent currents
of the system with repulsive interactions are larger than those of
the system with attractive interactions.
This counter intuitive result is due to the fact that local density
fluctuations are reduced in the presence of repulsive interactions.
\end{abstract}
\pacs{Pacs }
\narrowtext

\section{Introduction}

A hallmark of mesoscopic systems is the presence of permanent current
upon application of an external flux \cite{buttiker_permanent_current}.
Although a noninteracting electron theory is quite
successful in describing qualitatively the features of such an effect,
it strongly underestimates the magnitude of the observed
permanent currents compared to the
observed experimental values
\cite{levy_mesure_current,chandrasekhar_single_ring}.
One possible way out of this
discrepancy would be to include the electron-electron interactions.
In general the combined study of disorder and interactions proves
difficult to tackle, so that one has to resort to various
approximations. In three dimensions Hartree-Fock-like calculations
\cite{ambegaokar_interactions_current} tend to
suggest that permanent currents are enhanced by the interactions.
Arguments based on level statistics reach the same
conclusions
\cite{muller-groeling_spinless_current,muller-groeling_spinless_2d}.
In one dimension, on the other hand, numerical and analytical
techniques alike allow to treat the interactions exactly enough so that
firm results can be obtained.
Surprisingly, however, various independent
calculations performed on a spinless Hubbard
model with nearest neighbor interactions
\cite{bouzerar_spinless_currents,bouzerar_rg_current} or
long range coulomb interactions \cite{berkovits_coulomb_current}
have reached the opposite
conclusion, namely that repulsive interactions are detrimental to
permanent currents, which, therefore, casts doubts on the validity of
the approximate calculations in higher dimensions.

We here consider the full problem of interacting electrons
with spin, in the presence of disorder, and examine the permanent
currents in such a system. We will mainly concentrate for the
sake of simplicity
on a purely local interaction (Hubbard model), but will also consider
briefly finite range interactions.
The interactions are treated exactly and we use a
renormalization group calculation
\cite{giamarchi_loc_lettre,giamarchi_loc}
to take care of the disorder. We show
analytically that the permanent currents are {\bf enhanced} by the
presence of repulsive interactions, and that the result of the spinless
model, although correct, was an artifact.
As a byproduct of the study we also give a very simple formulation of
the Bethe-Ansatz equations for the pure attractive Hubbard model.

The plan of the paper is as follows. Section~\ref{Meissner}
relates the so called charge stiffness to the permanent currents, and
discusses various peculiarities that can occur for finite temperature.
Section~\ref{shubbard} introduces the model and treats
the interactions using the bosonization procedure. This section is
merely to fix the notations. Section~\ref{currents} computes the
stiffness using a renormalization group calculation.
Section~\ref{bansatz} focuses on the case of attractive interactions,
both with a large $-U$ expansion and using Bethe ansatz, and
section~\ref{longrange} examines an extended (nearest neighbors)
Hubbard model. Finally the
conclusions of our study can be found in section~\ref{conclusions}.

\section{Stiffness versus Meissner}
\label{Meissner}

Instead of computing the permanent currents it is simpler to
focus on the so-called conductivity stiffness $D$\cite{drudedudes}, which
measures the strength of the Drude peak in a macroscopic system
$\sigma(\omega) = D \delta(\omega) + \sigma_{\rm reg}$.
The stiffness $D$ can be related to the change of the energy of the
ground state of the system in presence of an external flux by
\begin{equation}
D = \frac{L}2 \left. \frac{d^2 E_0}{d \phi^2} \right|_{\phi=0},
\end{equation}
$E_0$ being the ground state energy of a ring in a field. Here $\phi$
denotes the boundary angle $\phi = 2 \pi f / f_0$ where $f$ is the flux
threading the ring and $f_0 = h c/e$ is the flux quantum.
On the other hand, for a mesoscopic system, the permanent current
measures the response to a finite flux by
\begin{equation}
J = L \left. \frac{d E_0}{d \phi} \right|_{\phi}
\end{equation}
Therefore the stiffness $D$ provides a measure of the permanent currents
for small (or close to a multiple of $2\pi$) flux since
$J =  2 D \phi$. Although the complete calculation of the permanent
currents at finite flux is also possible for a one dimensional
interacting system, the calculation is more complicated in the presence
of
disorder, and the stiffness carries enough information for our
present purposes.

In order to compute the permanent currents, one should in
principle distinguish
between an odd and even number of electrons, assumed
spinless for the present argument. For an odd number of
electrons the energy is minimum in zero external flux, and the number of
right and left moving electrons is equal. For an even number of
electrons, due to the $k=0$ state, the number of right and left moving
electrons cannot be equal, and the energy is not minimum at zero flux.
If one has non-disordered noninteracting electrons, it is obvious that
in that case the minimum of the energy corresponds to half a flux
quantum. An external flux can be incorporated in the wave function by
making the usual gauge transformation
\begin{equation}
c_i \to e^{i A r_i} c_i
\end{equation}
where $A$ is the vector potential.
This transformation modifies the kinetic energy term and absorbs an
external flux at the cost of a twist in boundary conditions. Since
this transformation is purely local and the interactions and disorder
do depend on density only, it will not affect the interaction and
disorder terms. Therefore the permanent currents for a system with an
even number of electrons, even in the presence of disorder and
interactions, will be identical to those of a system with an odd number
of electrons, with a shift corresponding to half a flux quantum.
In the calculation of
the stiffness one implicitly assumes that the equilibrium state is a
minimum of energy with zero current (equal number of right and left
carriers). Therefore
the stiffness for the case of an even number of electrons,
measures the permanent current
produced as a function of $(\phi-\phi_0/2)$.

The above connection between the permanent currents and the stiffness is
valid only at zero temperature. At finite temperature the permanent
current is given by the derivative of the free energy.
One could think that the stiffness would be given by the
second derivative of the free energy, within the canonical ensemble.
But such a quantity is
the so called Meissner fraction
\begin{equation}
\rho_s = \frac{L}2 \frac{d^2 F}{d \phi^2}
\end{equation}
Although both of $\rho_s$ and $D$ are
related to the current correlation function, they correspond to
different limits. $\rho_s$, being a thermodynamic quantity corresponds
to the limit $\omega\to 0$ first and then to $q\to 0$, whereas $D$ which
is related to transport corresponds to the limits taken in the reverse
order. For finite $T$ the two quantities are distinct.
For a macroscopic system $\rho_s$ measures the
superfluid density and
will be zero for a non-superfluid system, whereas $D$ can be non zero
if the system is perfectly conducting but not superconducting (e.g.
free electrons in the absence of impurities).
If one has a finite system then $\rho_s$ needs not to be zero even
if the system is not superconducting. In general the Meissner fraction
is given by
\begin{eqnarray} \label{stiff}
\rho_s & = & \frac{L}2 \left. \frac{d^2 F}{d \phi^2} \right|_{\phi=0} =
\sum_n \langle n | \frac{\partial^2 H}{\partial \phi^2}|n\rangle
e^{-\beta E_n} + \sum_{\alpha,\gamma,E_\alpha \ne E_\gamma}
\frac{e^{-\beta E_\alpha}-e^{-\beta E_\gamma}}{E_\alpha-E_\gamma}
|\langle \alpha | \frac{\partial H}{\partial \phi}|\gamma \rangle|^2 \\
& & - \beta  \sum_{\alpha,\gamma,E_\alpha = E_\gamma}
|\langle \alpha | \frac{\partial H}{\partial \phi}|\gamma \rangle|^2
e^{-\beta E_\alpha} \nonumber
\end{eqnarray}
where $n$ denotes a state of the system, $H$ is the Hamiltonian and
$E_n$ the energy of the eigenstates. We have separated out the contribution
from the manifold of degenerate states, which are of especial importance
for charge transport.
 In (\ref{stiff}) the limits
$L\to \infty$ and $T\to 0$
do not commute in general. If one takes the limit $L\to \infty$ first,
then $\rho_s$ tends to zero unless the system is really superconducting,
as can be checked by computing explicitly $\rho_s$ for e.g. free
electrons. For a mesoscopic system, where $L$ is finite, $\rho_s$ will
be finite and gives the slope of the permanent currents with the
flux
$J \sim \rho_s \phi$. One now has to distinguish whether we have an
odd or even number of electrons, or more generally whether the ground
state
is not degenerate (odd number of electrons) or whether it has a twofold
degeneracy
(even number of electrons). In the first case, for small enough
temperatures, the sums in (\ref{stiff}) will be dominated by the ground
state, other terms being suppressed by factors like $e^{-\beta
\Delta}$, where $\Delta$ would be the gap between the ground state and
the first excited states. Such a gap remains finite for a systems of
finite size. If the ground state is not degenerate then the average
value of the current in zero external flux is zero in the ground state.
One has
\begin{equation}
\langle \phi | \frac{\partial H}{\partial \phi}| \phi \rangle =
\langle \phi | J | \phi \rangle = 0
\end{equation}
The last term in (\ref{stiff}) vanishes and one finds a positive
$\rho_s$, roughly temperature independent (dominated by the first term
in (\ref{stiff})). There is, therefore, a paramagnetic permanent current
for small flux.

On the other hand, if the ground state has a twofold degeneracy, which
occurs e.g. for the case of an even number of electrons, then in each of
the degenerate ground states $\phi_\nu$ the current can (and will in
general) be nonzero
\begin{equation}
\langle \phi_\nu | \frac{\partial H}{\partial \phi}| \phi_\nu \rangle =
\langle \phi_\nu | J | \phi_\nu \rangle \ne 0
\end{equation}
In that
case (\ref{stiff}) is dominated by the last term, which gives rise to a
Curie type behavior, $\rho_s \sim - 1/T$. There will therefore be a
diamagnetic permanent current, with a slope proportional to $1/T$.

Many of  these properties are well known for free electrons (see e.g.
\cite{trivedi_conductivite_kubo} and references therein)
but the arguments
presented here show that they are much more general and rest only on
the degeneracy of the ground state and are valid
for interacting electrons as well.

\section{One dimensional Hubbard model: notations}
\label{shubbard}

Only a short derivation will be given here in order to fix the
notations. More details can be found in
\cite{solyom_revue_1d,emery_revue_1d}.

Let us consider for example the discrete 1D Hubbard model with $L$ sites
\begin{equation} \label{hubbard}
H = -t\sum_{\langle i,j,\sigma \rangle} c^\dagger_{i,\sigma} c_{j,\sigma} +
     U \sum_i n_{i,\uparrow} n_{i,\downarrow}
\end{equation}
where $\langle\rangle$ stands for nearest neighbors. Using the well
known boson representation of fermion operators in one dimension
the complete Hamiltonian (\ref{hubbard}) becomes, away from half-filling
\cite{solyom_revue_1d,emery_revue_1d}
\begin{equation} \label{complet}
H = H_\rho + H_\sigma  +
  \frac{2 g_{1\perp}}{(2\pi\alpha)^2}\int dx \cos(\sqrt8 \phi_\sigma(x))
\end{equation}
where $H_\rho$ and $H_\sigma$ are defined by
\begin{equation} \label{free}
H_\nu = \frac1{2\pi}\int dx \left[ (u_\nu K_\nu)(\pi \Pi_\nu)^2
         + (\frac{u_\nu}{K_\nu}) (\partial_x \phi_\nu)^2 \right]
\end{equation}
$\Pi$ and $\phi$ are canonically conjugate variables and
$\alpha$ is a short distance cutoff that can be identified with the
lattice spacing.
The $\rho$ and $\sigma$ parts of the Hamiltonian (\ref{complet})
respectively describe
the charge and spin degrees of freedom of the system.
The $g_{1\perp}$ term is the scattering between
electrons of opposite spins with an exchange of momentum of $2k_F$.
The Hamiltonian (\ref{complet}) describes in fact the most general one
dimensional Hamiltonian with spin conserving interactions,
provided that the proper values
of the $K$ and $u$ parameters are used. $K_\rho$ controls the charge
excitations. $K_\rho > 1$ corresponds to dominant superconducting
fluctuations, whereas $K_\rho < 1$ corresponds to charge- or spin-density
wave (depending on the spin part of the Hamiltonian) dominant fluctuations.

For the Hubbard Hamiltonian (\ref{hubbard}) the various
coefficients in (\ref{complet})
and (\ref{free}) are given, if $U$ is small compared to $v_F$, by
\begin{eqnarray} \label{param}
u_\rho K_\rho & = & u_\sigma K_\sigma = v_F \qquad,\qquad
u_\rho/K_\rho  =  v_F + U/\pi \\
u_\sigma/K_\sigma & = & v_F - U/\pi  \qquad,\qquad
g_{1\perp}  = U
\end{eqnarray}
For a macroscopic system $g_{1\perp}$ renormalizes to zero and
$K_\sigma$ to one for repulsive interactions.
In the asymptotic limit $u_\rho,u_\sigma,K_\rho$ are the only parameters
needed to describe the long range properties and have been computed
exactly using Bethe-Ansatz \cite{schulz_conductivite_1d}.
For attractive interactions there is a gap in the spin sector, and
only $u_\rho$ and $K_\rho$ are needed to describe the low energy
properties of the model. They can also be computed from
Bethe Ansatz \cite{kawakami_bethe_stiffness}
as will be seen in more details in section~\ref{bansatz}.

The charge stiffness $D$ can be obtained
\cite{schulz_conductivite_1d,shankar_spinless_conductivite}
from the bosonized Hamiltonian (\ref{complet}), and is simply
given by $D=2 u_\rho K_\rho$.

\section{Effect of disorder on $D(L)$}
\label{currents}

Disorder can be added to (\ref{complet}) by
\begin{equation} \label{dis}
H_{\rm dis} = \int dx W(x) \rho(x)
\end{equation}
where $W$ is a random potential. As is well known the forward scattering
due to the potential does not affect the conductivity in one dimension
(see e.g. \onlinecite{giamarchi_loc}) and
one can retain only the $2 k_F$ Fourier components of the random potential.
Higher Fourier components are less effective since they do not scatter
electrons on the Fermi surface, and do not correspond to low energy
processes. A notable exception are $4k_F$ components that will be
discussed later.
When expressed in term of the bosons variables (\ref{dis}) becomes
\begin{equation} \label{disbos}
H _{\rm dis} = \int dx \zeta(x) e^{i \sqrt2 \phi_\rho(x)}
                       \cos(\sqrt2 \phi_\sigma(x)) + \text{h.c.}
\end{equation}
where $\xi(x)$ is a complex random potential corresponding to the
part of the random potential having Fourier components close to $2 k_F$.
For simplicity we will take it Gaussian
\begin{equation}
\langle \zeta(x) \zeta^*(x')\rangle = W_\zeta \delta(x-x')
\end{equation}

The effect of (\ref{disbos}) can be computed using a renormalization group
calculation \cite{giamarchi_loc_lettre,giamarchi_loc}, where one changes
the cutoff (lattice spacing) $\alpha$ into $\alpha e^l$.
We will just
quote the results here
\begin{eqnarray} \label{renorm}
\frac{d K_\rho(l)}{dl} &=& -\frac12
\left(\frac{K_\rho^2u_\rho}{u_\sigma}\right) \Delta (l) \\
\frac{d K_\sigma (l)}{dl}  &=& - \frac12[\Delta(l)+ y(l)^2] K_\sigma ^2 \\
\frac{d u_\rho(l)}{dl} &=& -\frac{u_\rho^2K_\rho}{2u_\sigma} \Delta(l)\\
\frac{d u_\sigma (l)}{dl} &=& -\frac{u_\sigma   K_\sigma }{2}
                                                    \Delta(l)\\
\frac{d y(l)}{dl} &=& [2-2K_\sigma (l)]y(l) - \Delta (l) \\
\frac{d \Delta (l)}{dl}    &=& [3-K_\rho (l)-K_\sigma (l)-y(l)]
\Delta(l)
\label{renormfin}
\end{eqnarray}
with the dimensionless quantities:
\begin{equation}
\begin{array}{l@{\protect\hspace{2cm}}l}
\Delta = \frac{2W_\zeta\alpha}{\pi u_\sigma^2}
\left(\frac{u_\sigma}{u_\rho}\right)^{K_\rho}  &
y = g_{1\perp}/(\pi u_\sigma)
\end{array}
\end{equation}
The renormalizations of $K_\nu$ and $u_\nu$ are of first order in
$\Delta$ and $y^2$, and consequently can be neglected on the right sides
of the first three equations.
Contrary to the pure case, charge and spin degrees of freedom
are now no more decoupled.

For a macroscopic system, the physics implied by the equations
(\ref{renorm}-\ref{renormfin}) has been studied at length
\cite{giamarchi_loc_lettre,giamarchi_loc}. As can be seen from
(\ref{renormfin}), there are two regimes depending on the initial
values of $K_\rho,K_\sigma,y$. $\Delta$ can scale to zero, the
system is in that case delocalized, and as shown in
\cite{giamarchi_loc_lettre,giamarchi_loc} is dominated by divergent
superconducting fluctuations. In the other regime $\Delta$ scales
to large values, and the corresponding phase can be identified with a
localized phase. In that case the RG equations break down below a
certain length scale that can be identified with the localization
length. In the limit where $\Delta \to 0$, as can be seen from
(\ref{renormfin}) the localized-delocalized transition occurs when
$2-K_\rho = 0$ if $y >0$ (since $y$ renormalises to zero and $K_\sigma$
to one) or $3-K_\rho=0$ if $y<0$.
For a mesoscopic system of size $L$, one can expect the size to play
the role
of an infra-red cutoff in the RG equations. When the renormalized cutoff
$\alpha e^{l^*} \sim L$, i.e. $l^* = \log(L/\alpha)$,
one can treat the disorder term in perturbation, and
the stiffness is therefore simply given by the quadratic part of the
Hamiltonian, with the renormalized coefficients $D = 2 u(l^*) K(l^*)$.
Provided the size $L$ is
smaller than the length at which the equations cease to be valid one can
use them to
compute the various values of $K,u,y$ as function of the size $L$ of
the system, and from that to obtain the stiffness $D$.
Such a calculation is similar to the one performed for a
macroscopic system to get the exponent $K_\rho$ \cite{giamarchi_loc} at
finite temperatures. In that case the cutoff
is provided by the dephasing length $v_F/T$.

The full dependence of the stiffness on the size of the sample needs a
numerical integration of the RG equations, but the qualitative
features can be understood by looking at the very small disorder
limit. In that case one can neglect the renormalization of $u,K,y$ in
the equation for $\Delta$, which gives
\begin{equation}
\Delta = \Delta_0 e^{(3 - K_\rho - K_\sigma - g_{1\perp})\log(L/a)}
\end{equation}
where $\Delta_0$ is the initial value of the disorder.
Here we focus on the case of the Hubbard model. For small $U$
one can use the values of the parameters (\ref{param}) and one gets
\begin{equation} \label{debut}
\Delta = \Delta_0 (L/a)^{1-U/(\pi v_F)}
\end{equation}
Therefore the disorder grows more
slowly for the repulsive Hubbard model than in the attractive one.
In the same limit of a very small initial $\Delta$,
the stiffness is roughly given by
\begin{equation} \label{reduc}
D(l^*=\log(L/\alpha)) = D(l=0) - \text{Cste} \int_{l=0}^{l=l^*}\Delta(l)
\end{equation}
The bare stiffness $D(l=0)$ can be considered roughly independent of the
interactions if the system is far enough from half filling as can be
seen from (\ref{param}).
The dependence in $U$ comes only from lattice effects
\cite{shastry_twist_1d,schulz_conductivite_1d,giamarchi_umklapp_1d}
that breaks galilean invariance and are sensitive, for repulsive
interactions, only for fillings close to a
commensurate filling (mainly half filling where the pure system would
be a Mott insulator). For attractive
interactions the renormalization of the stiffness of the pure system due
to interactions will become much more important and will be
discussed in section~\ref{bansatz}. We will ignore in the
following the change of the bare stiffness due to the interactions and
will only focus on the effects due to the disorder. As can be seen from
(\ref{reduc}), the
disorder term tends to drastically reduce the stiffness $D$,
and this effect will be smaller
for the repulsive model than the attractive one and
the stiffness (the permanent currents) will be {\bf
enhanced} by repulsive interactions for a given size and a given
disorder.

For finite disorder one has to numerically integrate the RG equations.
The result is shown in figure~\ref{figure1} and is in agreement
with the simplified analysis above.
This rather counter-intuitive result can be simply understood with the
following argument: with the repulsive Hubbard model, the ground state
is almost a spin density wave (with a power-law decay of the
correlation functions) whose density is uniform. Such a ground
state couples
very weakly to non-magnetic impurities as is obvious from (\ref{dis}).
To couple to disorder, one has to
distort the spin density wave and make a fluctuation of the density, a
process that will cost an energy increasing with $U$. The disorder
effect is therefore very weak, at least if the size of the system is not
too big. On the other
hand the attractive Hubbard model has a ground state that contains
charge density wave fluctuations (although superconductive
fluctuations are the dominant ones) which can get very easily pinned
by impurities. On such a ground state the disorder will act very
efficiently and drastically reduce the stiffness compared to the pure
value therefore making the permanent currents smaller. Such an argument
is in agreement with higher dimensions
\cite{schmid_density_current}.

This is to be contrasted with
a previously studied spinless model
\cite{bouzerar_spinless_currents,bouzerar_rg_current,%
berkovits_coulomb_current}.
In that
case both the attractive and repulsive ground state have density
fluctuations, and both can be equally well pinned by disorder. Since in
the attractive case the superconducting fluctuations tend to screen the
disorder, the stiffness
increases with attractive interactions.
For the interactions to have a beneficent effect on the permanent
currents
one must necessary take a realistic model in which the main
effect of the interactions will be to homogenize the density as is
the case for the Hubbard model.

If the interactions are infinitely repulsive, the system becomes
equivalent to a model of spinless fermion with a Fermi momentum of
$2k_F$. In that case, although the $2k_F$ component of the disorder is
inefficient (as is also obvious from the fact that it will no more
correspond to a process on the new Fermi surface), one should worry
about the $4 k_F$ component of the disorder. Such a Fourier component
acts on the free spinless fermion, so that one recovers the stiffness
of free electrons in the presence of disorder. The crossover between the
two regimes would need a detailed analysis of the coupling of the $4k_F$
component of the charge density to disorder which is way beyond the
scope
of this paper, but one could naively expect a maximum of the permanent
currents for an intermediate value of the interactions.

If the size of the system becomes large enough the disorder will
renormalise to large values and the system will be localized. This is
always the case for repulsive interactions
\cite{apel_perturbation_localization,suzumura_scha_localization,%
suzumura_scha_localization_complet,giamarchi_loc_lettre,giamarchi_loc}.
For attractive interactions, a
localized-delocalized transition is in principle possible
\cite{apel_perturbation_localization,suzumura_scha_localization,%
suzumura_scha_localization_complet,giamarchi_loc_lettre,giamarchi_loc}
(for $K_\rho > 3$) and the stiffness could saturate to a finite value.
We will show in the following section that for the particular case of
the attractive Hubbard model, where one has only an on-site attraction,
this transition does not occur and the
system remains always localized.

\section{Negative $U$}
\label{bansatz}

Let us consider the case of a large negative $U$. In that case one would
naively imagine that the system should delocalize.
In fact a very large on-site
attraction cannot
delocalize, and {\bf increases} the localization (and therefore
decreases the permanent currents).
In the $U\to -\infty$ limit, one can perform
a large $|U|$ expansion. Only pairs of particles can
hop and if one introduces the operators $b_i = c_{i,\uparrow}
c_{i,\downarrow}$, the attractive Hubbard model then becomes a model
of hard core bosons with
a hopping $t'= t^2/|U|$ and a disorder $\Delta' = \Delta$. The
residual interaction between the bosons is also on the scale of
$t^2/U$. In fact using the
superexchange formulation or degenerate perturbation theory,
the model maps on precisely to the 1-d Heisenberg
antiferromagnet at a fixed magnetization (related to the density
of particles), with an exchange energy $4 t^2/U$.
By a Jordan Wigner transformation this
model corresponds to spinless fermion with a narrow bandwidth
and with nearest neighbors interaction, in the presence of the old
disorder.
Such a system is obviously localized, and since the kinetic energy
reduces with $|U|$ one expects the localization length to diminish when
the attraction is increased.

One can make the statements more quantitative for finite $U$
by studying the RG
equations in the attractive regime. In that case it is well known that
there is a gap in the spin excitation spectrum and that only the
charge sector remains ungapped. Keeping only the charge excitations
into account the RG equations become
\cite{giamarchi_loc_lettre,giamarchi_loc}
\begin{eqnarray} \label{renormatt}
\frac{d K_\rho(l)}{dl} &=& -\frac12 K_\rho^2 \Delta(l) \\
\frac{d u_\rho(l)}{dl} &=& -\frac{u_\rho^2K_\rho}{2} \Delta(l) \\
\frac{\Delta (l)}{dl}    &=& [3-K_\rho (l)]\Delta (l) \label{renormattf}
\end{eqnarray}
With
\begin{equation} \label{dimensionless}
\Delta(l) = (2 C_\sigma W_\xi \alpha)/(\pi u_\rho^2)
\end{equation}
and
$C_\sigma$ is a constant of order unity coming from the $\phi_\sigma$
correlations in the perturbation expansion. The equations (\ref{renorm})
can be used at scales above the size $\xi$ of a Cooper pair.
This approach will therefore be adapted for reasonably large
$U$. For small $U$ it will be better to use the equations
(\ref{renorm}-\ref{renormfin}), the crossover between the two regimes
occurring when $y \sim 1$.

In order to get the stiffness of the disordered system
one needs the initial values of $K_\rho$ and
$u_\rho$ in the absence of disorder as a function of the attraction
$U$. As for the repulsive Hubbard model, they can be deduced from the
Bethe-Ansatz solution \cite{kawakami_bethe_stiffness}.
We will here give a derivation based on an appealing
formulation introduced by Sutherland (in the absence of $\phi$)
involving the
formation of Cooper pairs which scatter without diffraction.
Let the number of particles be $2 M$, for which case
we note the Bethe Equations for the attractive $U$ Hubbard model,
with energy $E= -4 \sum_{j=1,M}  \cos(P_j/2 ) \cosh(Q_j)$, where
$(P_j,Q_j)=(Re,Im)\arcsin(\psi_j + i \frac{U}{4})$,
and $\psi_j$ satisfy the Bethe Equations \cite{sutherland_attractif_bethe}
\begin{equation}
L P(\psi_j)/2 = 2 \pi J_j + 2 \phi + \sum_{i=1,M}
\arctan[\frac{2}{U}(\psi_i-\psi_j)]
\end{equation}
where $J_j$ are integers (half odd integers) for M odd (even).
The flux
$\phi$ comes in with a factor of $2$ due to the charge of the Cooper pair.
These equations can also be obtained from the repulsive case
\cite{lieb_hubbard_exact,shastry_twist_1d}
by using
a particle hole transformation on the up electrons of a Half filled
model, the spin excitations then map on to the above equations. This can
be checked explicitly, using the idea of complementary
solutions due to Woynarovich \cite{woynarovich_transformation},
which essentially rests on the recognition
that the equations for real $k's$ in the Bethe equations of the
repulsive Hubbard model
are the $L$ real zeroes of a polynomial of degree $L + 2 M$. Hence
the residue theorem of Cauchy helps in transforming equations involving
the real $k's$ to those over complex $k's$. The complex $k's$ come in pairs,
and are essentially pinned to be $\psi_j \pm i \frac{U}{4}$,
in order to satisfy the growth conditions. The error involved
in writing down
the above Cooper pair representation is of $O(exp(-L/\xi(U))$, with
$\xi(U)$ the Cooper pair radius.

The parameters $u_\rho$ and $K_\rho$ can be
obtained by computing the compressibility $\chi =
u_\rho/K_\rho$ and the charge stiffness $D = 2 u_\rho K_\rho$ of the
pure system from the Bethe ansatz ground state energy.
Various values of $u_\rho$ and $K_\rho$ are plotted in
figure~\ref{figure2}
together with the stiffness $D$ (for the pure system).
As can be guessed from the large $U$ expansion
$u_\rho \to 0$ at large $U$. The
fact that the parameter $K_\rho$ remains finite shows that the
system remains interacting. One can check that the limiting value of
$K_\rho$
is in agreement with the one obtained for the $XXZ$ chain
\cite{haldane_xxzchain} on which this
system maps in the large $U$ limit. Due to the reduction of the
velocity,
the stiffness of the clean system itself goes to zero at large $U$.
Conversely to what happened for repulsive interactions where the
stiffness of the pure system was nearly interactions independent,
there is here a drastic {\bf reduction} of the stiffness when
the attraction is increased \cite{kawakami_bethe_stiffness}.

Using equations (\ref{renorm}) one computes the stiffness in the
presence of
disorder. Here the main in the decrease of the velocity
$u_\rho$, which increases the relative strength of the disorder given by
the dimensionless parameter (\ref{dimensionless}).
Some results are shown in figure~\ref{figure3}, where we have
normalized the stiffness to its value in the absence of disorder
to avoid the trivial effect of renormalization of the bare stiffness by
attractive interactions. In agreement with the previous section the
reduction of the stiffness due to disorder becomes more and more
important as the attraction $U$ is increased.

\section{Extended Hubbard model}
\label{longrange}

In order to check the validity of the arguments presented here for a
slightly more general model than the Hubbard model, we also look at an
extended Hubbard model with a nearest neighbor interaction
$V$ defined by
\begin{equation}
V \sum_i n_i n_{i+1}
\end{equation}
In that case, and for small $U$ and $V$ the various parameters entering
the equations are
\begin{eqnarray}
K_\rho  & \simeq &
        1 - \frac{U}{2\pi v_F} - \frac{V}{\pi v_F}(2 - \cos(2k_F a)) \\
K_\sigma  & \simeq &     \label{ksigma}
        1 + \frac{U}{2\pi v_F} + \frac{V}{\pi v_F}\cos(2k_F a) \\
y_\perp & \simeq & \frac{U}{\pi v_F} + \frac{V}{\pi v_F}\cos(2k_F a)
\end{eqnarray}
For small $U$ and $V$ the renormalization of velocities is of second
order in $U,V$ and can be neglected. When replaced in equations
(\ref{renormfin}) one gets
\begin{equation}  \label{renorv}
(3-K_\rho-K_\sigma-y) = 1 - \frac{U}{2\pi v_F} +
                        \frac{2 V}{\pi v_F}(1 - 2\cos(2k_F a))
\end{equation}
If $U \gg V$  the results are unchanged compared to the case of the pure
Hubbard model. In order to check whether the physical ideas introduced
here on the increase of the permanent current due to repulsive
interactions are correct, or whether they are an artifact of the purely
local Hubbard model, one can consider the artificial limit
where $U=0$ and $V$ remains finite.
Note that the model does not boil down in that case to the spinless
fermion
model since the $V$ term still introduces interactions among opposite
spins.

In that case the effect of the
interaction depends on the filling. For low filling, a repulsive $V$
will tend to favor a spin density wave ground state again, whereas an
attractive one would tend to pair particles on neighboring sites, giving
a modulation of the density. For large fillings the situation changes:
a positive $V$ will now tend to favor two particles on the same site, to
avoid paying the repulsion, and therefore to give a charge density wave.
An attractive $V$ favoring two particles on neighboring
site will this time give a spin density wave. The change between a SDW
towards CDW ground state occurs when $K_\sigma=1$ and as can be seen
from (\ref{ksigma}) this will occur when $k_F=\pi/4$.
Added to this
is the competing effect that the more attractive $V$ we have, the more
there
are superconducting fluctuations in the system which tend to reduce the
disorder. Above quarter filling the two effects go hand in hand
and repulsive interactions are detrimental to the stiffness, whereas
below quarter filling the two effects will compete. From
(\ref{renorv}) one can
see that the point where a repulsive $V$ again becomes favorable to the
stiffness is $k_F = \pi/3$. Below this filling, the fluctuations of the
density generated by a repulsive $V$ are too strong to be balanced by
the superconducting fluctuations and a positive $V$ will increase the
permanent currents.

\section{Conclusions}
\label{conclusions}

We have looked in this paper at the stiffness constant of a Hubbard
model as a function of the size of the system. The stiffness constant
is directly related to the permanent currents in the presence of an
external flux by $J = D \phi$ for small flux. We have shown that both
the attractive and repulsive Hubbard model are always localized for a
macroscopic system regardless of the strength of the interactions.
In fact, using Bethe Ansatz solution or for large $-U$ simple
perturbation theory, one shows that the localization length decreases
for attractive interactions due to the reduction of charge velocity.

For a mesoscopic system, the stiffness in the repulsive Hubbard model
is much less sensitive to disorder than for the attractive one.
Therefore the permanent currents are {\bf enhanced} by repulsive
interactions. This surprising result is
related to the fact that for the attractive Hubbard model the
ground state contains strong charge density wave fluctuations that
pins easily on the impurities, whereas repulsion favors a uniform
density and makes the pinning harder.
This property remains valid for a model with longer range interactions.
In general the effects of the interactions on the permanent currents
is controlled by two competing effects. One is the presence of density
fluctuations in the ground state. The more there will be, the more
easily the system will be pinned by disorder and the more the permanent
currents will be reduced compared to the pure value. In general
repulsive interactions will tend to favor a homogeneous density (local
fluctuations in density will cost an energy increasing with the
repulsion), and therefore will tend to increase the permanent currents. On
the other hand, attractive interactions promote superconducting
fluctuations in the system that tend to screen the disorder and
therefore tend to increase the permanent currents.

Previous studies of one
dimensional systems, leading to the conclusion that repulsive
interactions reduced permanent currents, were performed
on a spinless model. In such a rather artificial model,
the first effect does not occur, since density fluctuations are always
present both for attractive and repulsive interactions
and therefore repulsive interactions are detrimental to permanent
currents. In a more realistic model where the local interactions are
the dominant ones (interactions in a real system do decrease with
distance !), the density effect will dominate and the permanent current
are increased.

This study is, strictly speaking, restricted to one-dimensional
systems, and a direct comparison of our results with experimental,
three-dimensional, rings is not feasible. It nevertheless
suggests that in mesoscopic systems the presence of
repulsive interactions can considerably enhance the permanent
currents, and confirms in the exactly solvable one-dimensional case,
that
the increase of the permanent currents is linked to a reduction of
the local density fluctuations by the repulsive interactions.
It is therefore tempting to ascribe the discrepancies
observed between the measured and the computed (with a free electron
theory) values of the permanent currents to such an interaction
effect, an interpretation compatible with recent perturbative
calculations \cite{ramin_preprint}.

\section*{Acknowledgements}
It is a pleasure to thank G. Montambaux, H. Bouchiat and B. Reulet for
many illuminating discussions.


\begin{thebibliography}{10}

\bibitem{junk}
Laboratoire associ\'e au CNRS. email: giam@lps.u-psud.fr.

\bibitem{junk2}
bss@physics.iisc.ernet.in

\bibitem{buttiker_permanent_current}
M. B{\"u}ttiker, Y. Imry, and R. Landauer, Phys. Lett. A {\bf 96},  365
  (1983).

\bibitem{levy_mesure_current}
L.~P. Levy, G. Dolan, J. Dunsmuir, and H. Bouchiat, Phys. Rev. Lett. {\bf 64},
  2074  (1990).

\bibitem{chandrasekhar_single_ring}
V. Chandrasekhar {\it et~al.}, Phys. Rev. Lett. {\bf 67},  3578  (1991).

\bibitem{ambegaokar_interactions_current}
V. Ambegaokar and U. Eckern, Phys. Rev. Lett. {\bf 65},  381  (1990).

\bibitem{muller-groeling_spinless_current}
A. M{\"u}ller-Groeling, H.~A. Weidenm{\"u}ller, and C.~H. Lewenkopf, Europhys.
  Lett. {\bf 22},  193  (1993).

\bibitem{muller-groeling_spinless_2d}
A. M{\"u}ller-Groeling and H.~A. Weidenm{\"u}ller, Phys. Rev. B {\bf 49},  4752
   (1994).

\bibitem{bouzerar_spinless_currents}
G. Bouzerar, D. Poilblanc, and G. Montambaux, Phys. Rev. B {\bf 49},  8258
  (1994).

\bibitem{bouzerar_rg_current}
G. Bouzerar and D. Poilblanc preprint (1994).

\bibitem{berkovits_coulomb_current}
R. Berkovits, Phys. Rev. B {\bf 48},  14381  (1993).

\bibitem{giamarchi_loc_lettre}
T. Giamarchi and H. Schulz, Europhys. Lett. {\bf 3},  1287  (1987).

\bibitem{giamarchi_loc}
T. Giamarchi and H.~J. Schulz, Phys. Rev. B {\bf 37},  325  (1988).
\bibitem{drudedudes}
W. Kohn, Phys. Rev. {\bf 133}, A171 (1964),  Ref\cite{shastry_twist_1d},
D. J. Scalapino, S. R. White and S.C. Zhang, Phys. Rev. Lett. { \bf 68},
2830 (1992).

\bibitem{trivedi_conductivite_kubo}
N. Trivedi and D. Browne, Phys. Rev. B {\bf 38},  9581  (1988).

\bibitem{solyom_revue_1d}
J. S\'olyom, Adv. Phys. {\bf 28},  209  (1979).

\bibitem{emery_revue_1d}
V.~J. Emery,  in {\em Highly Conducting One-Dimensional Solids}, edited by
  J.~T.~D. et~al. (Plenum, New York, 1979), p.\ 327.

\bibitem{schulz_conductivite_1d}
H. Schulz, Phys. Rev. Lett. {\bf 64},  2831  (1990).

\bibitem{kawakami_bethe_stiffness}
N. Kawakami and S.~K. Yang, Phys. Rev. B {\bf 44},  7844  (1991).

\bibitem{shankar_spinless_conductivite}
R. Shankar, Int. J. of Modern Physics B {\bf 4},  2371  (1990).

\bibitem{shastry_twist_1d}
B. Shastry and B. Sutherland, Phys. Rev. Lett. {\bf 65},  243  (1990).

\bibitem{giamarchi_umklapp_1d}
T. Giamarchi, Phys. Rev. B {\bf 44},  2905  (1991).

\bibitem{schmid_density_current}
A. Schmid, Phys. Rev. Lett. {\bf 66},  80  (1991).

\bibitem{apel_perturbation_localization}
W. Apel and T.~M. Rice, Phys. Rev. B {\bf 26},  7063  (1982).

\bibitem{suzumura_scha_localization}
Y. Suzumura and H. Fukuyama, J. Phys. Soc. Jpn. {\bf 52},  2870  (1983).

\bibitem{suzumura_scha_localization_complet}
Y. Suzumura and H. Fukuyama, J. Phys. Soc. Jpn. {\bf 53},  3918  (1984).

\bibitem{sutherland_attractif_bethe}
B. Sutherland,  in {\em Vol 242 Exactly Solvable Problems in Condensed
Matter and Relativistic Field Theory}
, edited by B. S. Shastry,
V. Singh and S.S. Jha
(Springer, New York, 1985).

\bibitem{lieb_hubbard_exact}
E.~H. Lieb and F.~Y. Wu, Phys. Rev. Lett. {\bf 20},  1445  (1968).

\bibitem{woynarovich_transformation}
F. Woynarovich,  J. Phys. C {\bf 16},  6593  (1983).

\bibitem{haldane_xxzchain}
F.~D.~M. Haldane, Phys. Rev. Lett. {\bf 45},  1358  (1980).

\bibitem{ramin_preprint}
M. Ramin, B. Reulet and H. Bouchiat, preprint (1994).

\end{thebibliography}

\newpage
\begin{figure}
\caption{Normalized
stiffness $D/D_0$
as a function of the size of the system (in units
of the lattice spacing $\alpha$) obtained by numerically integrating the
RG equations (\protect{\ref{renorm}}-\protect{\ref{renormfin}}). $D_0$
is the stiffness in the absence of disorder. All energies are in units
of the orignial Fermi velocity $v_F$.
The disorder $W_\xi/v_F$ is fixed to $W/v_F = 5 \; 10^{-4}$.
The full line is $U/v_F=0$, the dotted line $U/v_F=-0.5$ and the
dash-dotted line $U/v_F=0.5$. For a given size $L$, systems
with repulsive
interactions have a larger stiffness than those with attractive ones.
\label{figure1}}
\end{figure}
\begin{figure}
\caption{Values of $u_\rho$ and $K_\rho$ for the attractive Hubbard
model as a function of the strength of the interaction $|U|$. These
values are obtained by numerical integration of the Bethe-Ansatz
equations for systems of $L=200$ sites with respectively $90$, $70$,
$50$ particles per spin for the full, dotted and dash-dotted lines.
This corresponds to density of $n=0.9$, $n=0.7$ and $n=0.5$ particles
per sites respectively.
\label{figure2}}
\end{figure}
\begin{figure}
\caption{Normalized
stiffness $D/D_0$ for the attractive Hubbard model
as a function of the size of the system (in units
of the lattice spacing $\alpha$) obtained by numerically integrating the
RG equations (\protect{\ref{renormatt}}-\protect{\ref{renormattf}}).
$D_0$ is the stiffness in the absence of disorder.
The effective disorder $C_\sigma W_\xi$ is fixed to
$C_\sigma W_\xi = 5 \; 10^{-4}$ and the density is $n=0.5$ particles per
site.
The full, dotted and dash-dotted lines correspond respectively to
$|U|=5$, $|U|=10$ and $|U|=15$. The corresponding bare stiffness are
respectively $D_0=1.78$, $D_0=1.04$ and $D_0=0.72$.
Here again, for a given size $L$ and fixed disorder the stiffness
decreases with increasing attraction.
\label{figure3}}
\end{figure}
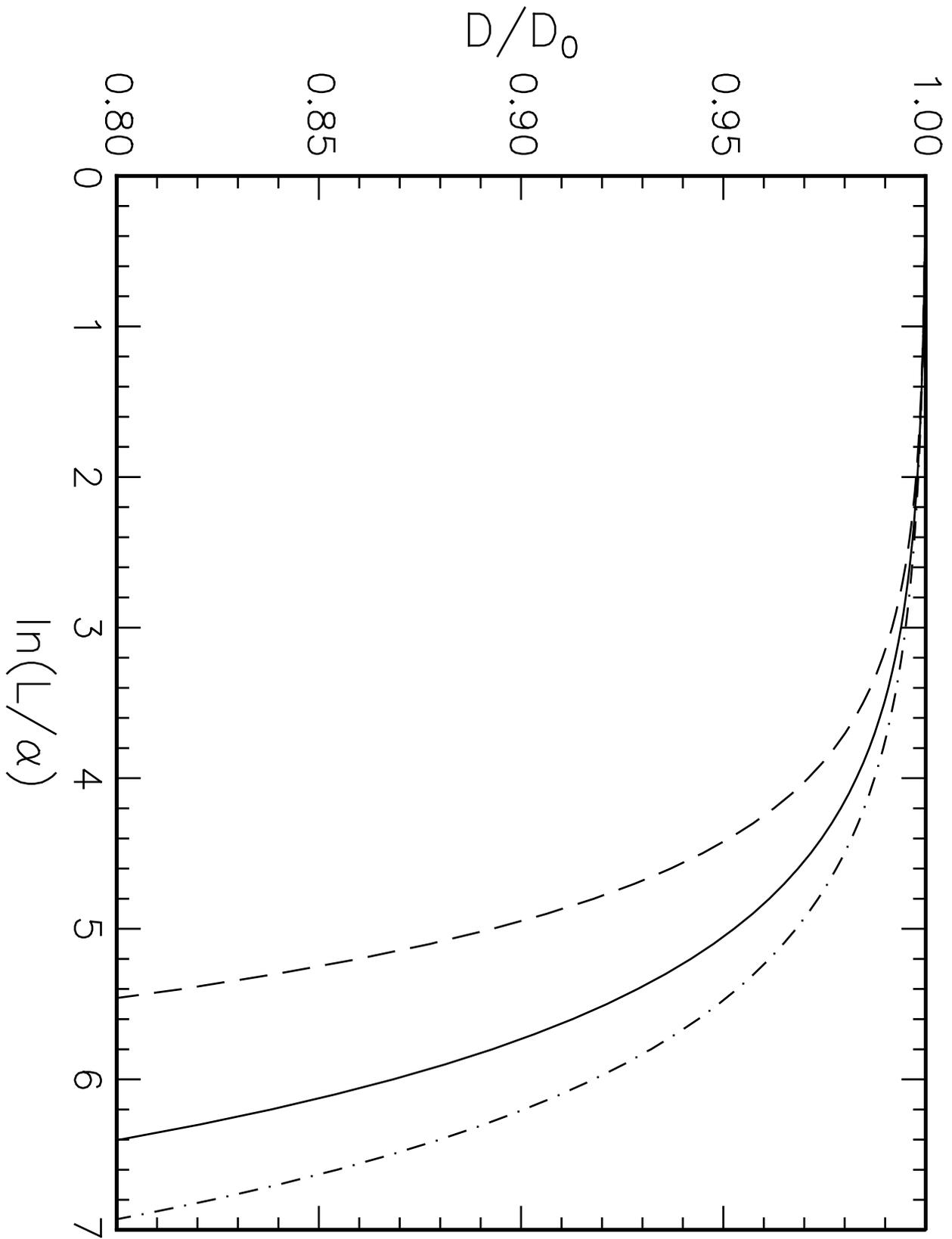
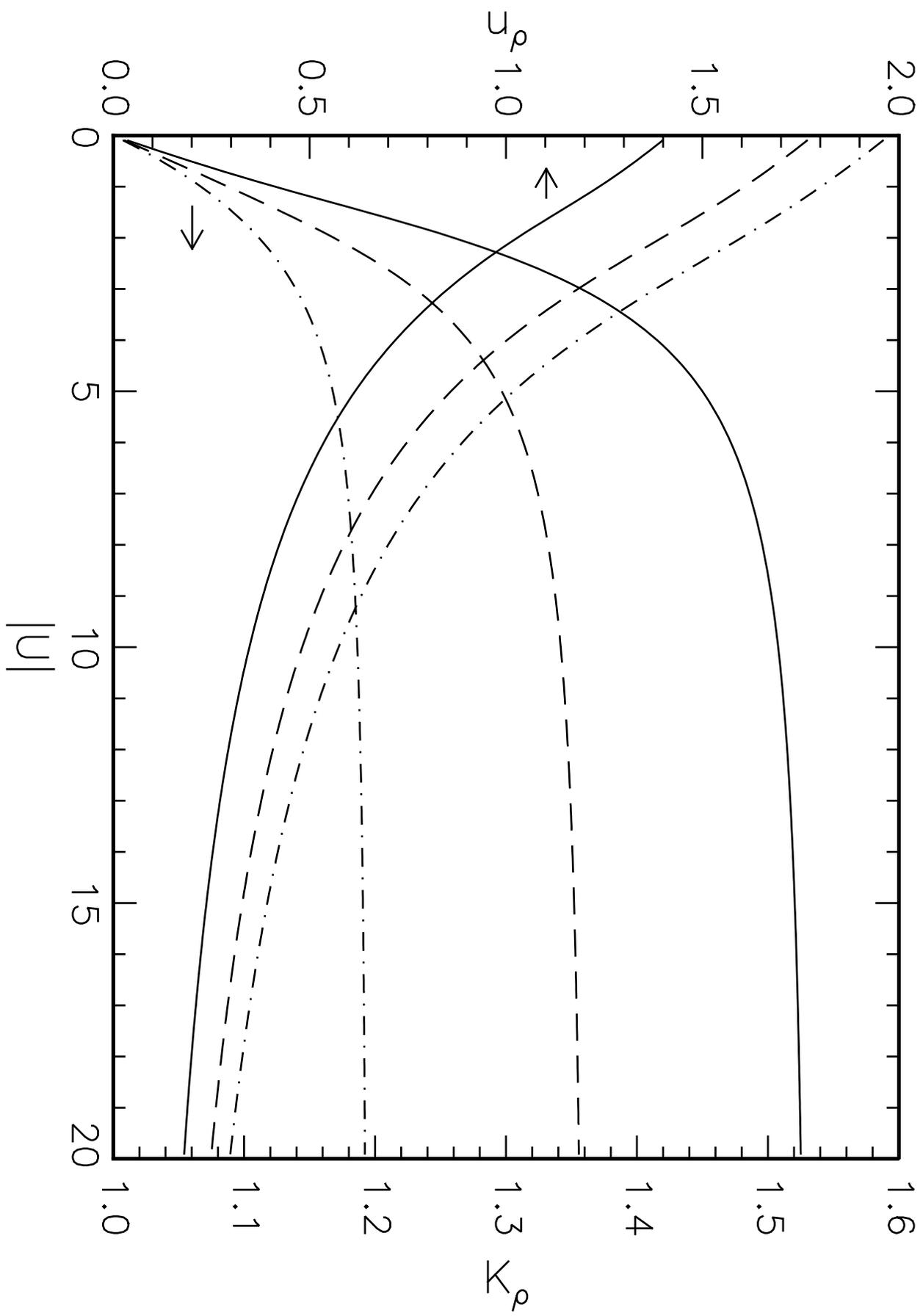
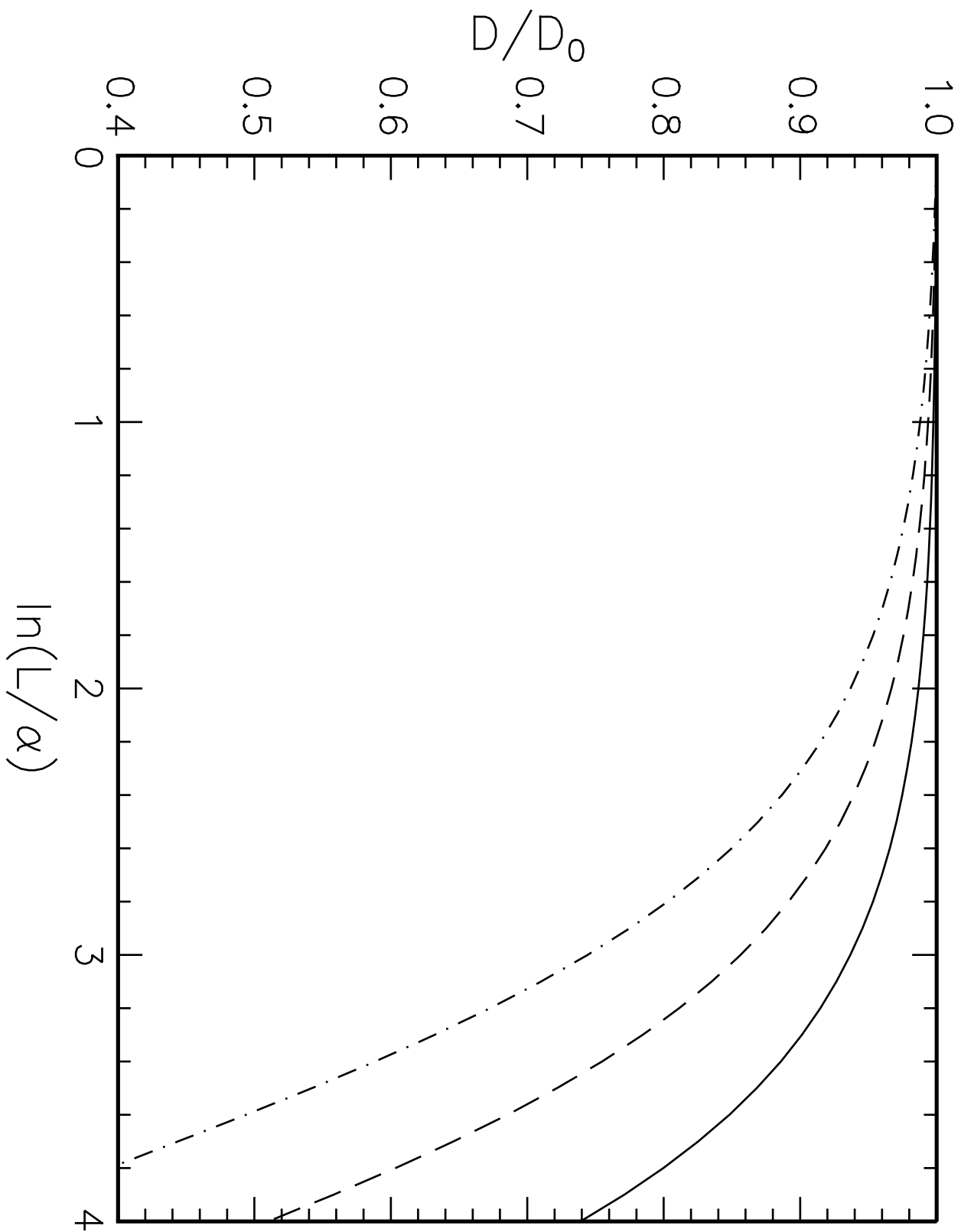
\end{document}